\documentclass[12pt]{article}
\usepackage[english]{babel}
\usepackage{amsmath}
\usepackage{amssymb}
\usepackage{bbm}
\usepackage{color}
\usepackage[titletoc,toc,title]{appendix}
\Roman{section}
\newcommand{\naw}[1]{\left(#1\right)}

\newcommand{\com}[1]{\left[#1\right]}
\newcommand{\modu}[1]{\left|#1\right|}
\newcommand{\poisson}[1]{\left\{#1\right\}}

\begin{document}

\begin{center}
\textsc{\Large{Note on G\"uney-Hillery approach to Bell inequalities}}
\newline

\large{Katarzyna Bolonek-Laso\'n}\footnote{katarzyna.bolonek@uni.lodz.pl}\\ 
\emph{\normalsize{Faculty of Economics and Sociology, Department of Statistical Methods, \\University of Lodz,
41/43 Rewolucji 1905 St., 90-214 Lodz,  Poland.}}\\
\large{Piotr Kosi\'nski}\footnote{piotr.kosinski@uni.lodz.pl}\\
\emph{\normalsize{Department of Computer Science, Faculty of Physics and Applied Informatics, University of Lodz, 149/153 Pomorska St., 90-236 Lodz, Poland.}}
\end{center}

\begin{abstract}

We analyze certain aspects of group theoretical approach to Bell inequalities proposed by G\"uney and Hillery. The general procedure for constructing the relevant group orbits is described. Using Hall theorem we determine the form of Bell inequality in the single orbit case. It is shown that in this case the Bell inequality is not violated for maximally entangled state generating trivial subrepresentation if the representation under consideration is real. 
\end{abstract}

\section{Introduction}
In his groundbreaking paper Bell \cite{Bell} showed that any realistic theory obeying locality condition must satisfy certain conditions which can be expressed in terms of inequalities known as Bell inequalities. Their actual form depends on the context, in particular the number of parties, measurement settings and possible outcomes for each measurement \cite{Clauser}$\div$\cite{Cabello}; excellent reviews are provided in \cite{Liang},\cite{Brunner}.

Bell inequalities are, in general, violated in the quantum mechanical case. This is their crucial property which allows, to some extent, to make the quantitative distinction between classical and quantum regimes.

Another notion which plays an important role in the description of physical phenomena is that of symmetry. Mathematically, symmetries are desrcibed by the relevant groups and their representations. A group can act as usual symmetry group or dynamical symmetry group (in the case the Hamiltonian does not belong to the center of universal enveloping algebra). The former classifies the states within the space of states while the latter generates the (minimal) space of states; this statement refers both to classical and quantum case.

In view of the role played by symmetries it appears desirable to analyze the Bell inequalities and their quantum violation in the context of physical systems exhibiting symmetries of various kinds. Such an analysis has been initiated in two nice papers by G\"uney and Hillery \cite{Guney}, \cite{Guney1} and continued in \cite{Bolonek}$\div$\cite{Yang}. 

The great advantage of the group theoretical approach to Bell inequalities is that one can find quite easily the upper quantum bound on the relevant sum of probabilities (the counterpart of Cirel'son bound \cite{Cirelson}) using elementary representation theory of finite groups. This upper bound should then be compared with that corresponding to classical probabilities, i.e. the one resulting from Bell inequality, to find out whether the latter is violated or not.

On the other hand, contrary to the quantum upper bound, the group theoretical meaning of the classical one is less understood. It is known \cite{Guney}$\div$\cite{Yang} that the possibility of violation of Bell inequalities depends crucially on the number and form of the orbits selected in the representation space of the group under consideration. It would be therefore desirable to shed some light on the problem of the choice of the appropriate set of orbits relating it to the particular structure of the relevant symmetry group.

The present paper initiates this line of research. First of all, we outline the general approach to the construction of group orbits in representation space which consist of a number of disjoint orthonormal bases defining the spectral decompositions of observables. We characterize such orbits in terms of cyclic subgroups and the corresponding cosets. The approach proposed here provides a generalization of the examples considered in Refs. \cite{Guney}$\div$\cite{Yang}. Then, using the Hall theorem, we show that it is rather unlikely to find reasonable examples of Bell inequalities violation if only a single group orbit is involved. To this end we consider real irreducible (over $\mathbb{C}$) representation of some finite group $G$. The tensor square of such representations contains trivial representation as subrepresentation. It is spanned by a maximally entangled vector. We generally expect that the more the state vector is entangled the stronger is the violation of Bell inequality. However, in the case under consideration the Bell inequality is not violated if only one orbit is involved. In fact, both classical and quantum bounds are equal and saturated. 

The example analyzed in the present paper is rather specific. It would be interesting to extend the analysis to more general situation of arbitrary (not necessarily real) representations and arbitrary vectors in the total of states. This will be the subject of subsequent publications.

Finally, let us note that various applications of group theory to Bell inequalities were already discussed by several authors. In \cite{Sliwa} a number of symmetries, like permutation of observables, were considered which allowed to split the set of all Bell inequalities into rbits consisting of equivalent inequalities. In \cite{Bancal} the notion of symmetric Bell inequalities has been introduced. Our approach differs basically as it is concentrated on group-theoretical construction of states defining the spectral decompostions of observables entering relevant inequalities.

The paper is organized as follows. In Sec.~II we describe the group-theoretical approach to Bell inequalities. Sec.~III is devoted to the construction of group orbits consisting of the number of orthonormal bases which provide spectral decompositions of some observables. In Sec.~IV we apply Hall theorem to find the (saturated) upper bound on  the sum of classical probabilities corresponding to a single group orbit. We also show that the maximally entangled state vector spanning trivial subrepresentation in the space of states of total system does not provide the violation of Bell inequality. Some conclusions are presented in Sec.~V. 

\section{Bell inequalities in group-theoretical setting}

As it is well known Bell inequalities provide a quantitative tool for answering the question whether the quantum mechanical predictions can be explained on classical level in terms of the so called hidden variables. One can set the following scheme for Bell inequalities. Consider a two-partite (Alice and Bob) physical system and pick the sets $\poisson{A_1,\ldots, A_n}$ and $\poisson{B_1,\ldots, B_n}$ of observables for Alice and Bob, respectively; in general, some observables may appear in the above collections more than once. We assume further that $\com{A_i, B_j}=0$ for all $1\leq i,j\leq n$; thus both on the classical and quantum levels the joint probabilities
\begin{equation}
p_{ij}\naw{a_i,b_j}=p\naw{A_i=a_i, B_j=b_j}\label{a}
\end{equation}
are well defined. The classical (hidden variables) case is distinguished by further assumption that all observables $A_1,\ldots, A_n, B_1,\ldots,B_n$ can be simultaneously measured. Therefore, we assume the existence of joint probability distribution
\begin{equation}
p\naw{\underline{a},\underline{b}}=p\naw{A_1=a_1,\ldots, A_n=a_n,B_1=b_1,\ldots,B_n=b_n}
\end{equation}  
from which the probabilities (\ref{a}) are returned as marginals,
\begin{equation}
p_{ij}\naw{a_i,b_j}=\sum_{\begin{subarray}{l}
\naw{\underline{\widetilde{a}},\underline{\widetilde{b}}}:\widetilde{a}_i=a_i\\
\qquad\widetilde{b}_i=b_i\end{subarray}}p\naw{\underline{\widetilde{a}},\underline{\widetilde{b}}}\label{a1}
\end{equation}
We are interested in estimating the sums of probabilities of the form 
\begin{equation}
S={\sum_{\naw{i,j}}{}}^\prime{\sum_{\naw{a_i,b_j}}}^\prime p_{ij}\naw{a_i,b_j}\label{a5}
\end{equation}
where the primes mean that the summations run over some subsets of the pairs $\naw{i,j}$ and $\naw{a_i,b_j}$.

In the classical case one can insert for $p_{ij}\naw{a_i,b_j}$ the right hand side of eq. (\ref{a1}) yielding
\begin{equation}
S=\sum_{\naw{\underline{\widetilde{a}},\underline{\widetilde{b}}}}c\naw{\underline{\widetilde{a}},\underline{\widetilde{b}}}p\naw{\underline{\widetilde{a}},\underline{\widetilde{b}}}\label{a2}
\end{equation}
where the sum runs over all joint configurations $\naw{\underline{\widetilde{a}},\underline{\widetilde{b}}}=\naw{\widetilde{a}_1,\ldots, \widetilde{a}_n,\widetilde{b}_1,\ldots,\widetilde{b}_n}$ while the nonnegative integers $c\naw{\underline{\widetilde{a}},\underline{\widetilde{b}}}$ indicate the number of times a particular joint configuration $\naw{\underline{\widetilde{a}},\underline{\widetilde{b}}}$ enters the sum defining $S$. Now, due to $0\leq p\naw{\underline{\widetilde{a}},\underline{\widetilde{b}}}\leq 1$, $\sum\limits_{\naw{\underline{\widetilde{a}},\underline{\widetilde{b}}}}p\naw{\underline{\widetilde{a}},\underline{\widetilde{b}}}=1$, eq. (\ref{a2}) implies
\begin{equation}
S\leq \max\limits_{\naw{\underline{\widetilde{a}},\underline{\widetilde{b}}}}c\naw{\underline{\widetilde{a}},\underline{\widetilde{b}}}\label{a3}
\end{equation}
which is the Bell inequality. Let us note that even if the maximum is attained by more than one $c\naw{\underline{\widetilde{a}},\underline{\widetilde{b}}}$, one can always select the joint probability maximizing $S$ in the form
\begin{equation}
p\naw{\underline{a},\underline{b}}=\left\{\begin{array}{ll}
1 & \text{if}\,\, \naw{\underline{a},\underline{b}}=\naw{\underline{a}^{\naw{0}},\underline{b}^{\naw{0}}}\\
0 & \text{otherwise.}
\end{array}\right.
\end{equation}

Bell inequality (\ref{a3}) can be converted into standard form once the probabilities $p_{ij}\naw{a_i,b_j}$ are expressed in terms of relevant correlation functions.

Let us compare the approach described above with the standard framework for studying Bell inequalities (for the detailed review of the latter see Ref. \cite{Brunner}). One considers the system consisting of two parties (Alice and Bob) representing distant observes; it should be stressed that the assumption that the observes are distant simply reduces to the statement that the Alice observables commute with those of Bob. Alice and Bob inputs, $x$ and $y$, are labeled by the elements of the set $\poisson{1,2,\ldots,m}$ while their outputs, $a$ and $b$ - by the elements of $\poisson{1,2,\ldots,\Delta}$. In our setting the Alice (Bob) inputs correspond to the choice of observables $A_i$ ($B_j$) and their outputs are the results of measurements $a_i$ ($b_j$). Now, in the probability space $\mathcal{P}\subset \mathbb{R}^{\Delta^2m^2}$ one distinguishes two subsets: local set $\mathcal{L}$, characterized by the existence of hidden variables representation and the no-signalling set $\mathcal{NS}$ consisting of probabilities obeying no-signalling conditions; one notes that $\mathcal{L}\subset\mathcal{NS}$ and both sets are polytopes. Bell inequalities define the hyperplanes delimiting $\mathcal{L}$. In the framework presented above our starting point is, following Refs. \cite{Guney}, \cite{Guney1}, the assumption that there exists a joint probability distribution for \underline{all} observables of $A$ and $B$; this the purely classical situation as on the quantum level the Alice (Bob) observables do not commute with each other and obey the appropriate uncertainty relations. With such an assumption one arrives at the inequality (\ref{a3}). As shown by Fine \cite{Fine}, \cite{Fine1} (see also \cite{Halliwell}, \cite{Halliwell1}) both approaches are equivalent, at least for $\Delta=2$, i.e. the inequalities (\ref{a3}) characterize $\mathcal{L}$. However, it must be stressed that this is the case provided the maximal set of independent inequalities (\ref{a3}) is considered. In our case we consider specific sums (\ref{a5}) resulting from group action in the space of states. In spite of that the breakdown of our inequalities is sufficient to conclude that the classical picture doesn't work. It corresponds to the statement that for a given point in $\mathcal{P}$ it is sufficient to show that it lies on the wrong side of some hyperplane delimiting $\mathcal{L}$ to conclude that it cannot belong to $\mathcal{L}$.

On the quantum level the sum $S$, eq. (\ref{a5}), takes the following form. Any observable $A_i$ $\naw{B_j}$ is represented by a selfadjoint operator acting in the Alice (Bob) space of states $\mathcal{H}$; the space of states of the bipartite system is simply $\mathcal{H}\otimes \mathcal{H}$. Denote by $\poisson{v^\mu_{a_i}}^{\dim\mathcal{H}}_{\mu=1}$ ($\poisson{v^\nu_{b_i}}^{\dim\mathcal{H}}_{\nu=1}$) the relevant eigenvectors
\begin{equation}
A_i v^\mu_{a_i}=a^\mu_i v^\mu_{a_i},\quad B_i v^\nu_{b_i}=b^\nu_i v^\nu_{b_i}
\end{equation}
and let, for any $w\in\mathcal{H}\otimes\mathcal{H}$
\begin{equation}
S={\sum_{\naw{i,j}}}^\prime{\sum_{\naw{\mu_i,\nu_j}}}^\prime \modu{\naw{v^{\mu_i}_{a_i}\otimes v^{\nu_j}_{b_j},w}}^2.\label{b}
\end{equation}
One can select the summations in eqs. (\ref{a5}) and (\ref{b}) in such a way that eq. (\ref{b}) becomes the quantum representation of (\ref{a5}).

Following \cite{Guney}, \cite{Guney1} (see also \cite{Bolonek}) we shall consider the special case when the operators and states under consideration are obtained through the action of some finite group $G$. As it has been explained in the Introduction this is motivated by the fact that the notion of symmetry plays the important role in the description of many (if not most) physical systems. On the formal level the symmetry transformations are decribed by the action of certain group. In the quantum case, due to the superposition priniciple, this action must be linear and we arrive at the linear unitary representation of the group under consideration. It is natural to consider a given physical system as consisting of a number of parties. We shall assume that this decomposition is invariant under the group action which implies that the group acts in the total space of states through the tensor product of representations acting in individual parties. On the other hand we can assume that the group action on each party is irreducible; in fact, for finite groups (as well as compact ones and many noncompact interesting from physical point of view) any representation decomposes into direct sum of irreducible ones. 

From physical point of view any state obtained from the initial one by group action is equally relevant (actually, physics is described in terms of the relations between such states). In this way we arrive at the notion of orbits in the space of states. The main idea of G\"uney and Hillery is to select some initial product vector $v_A\otimes v_B$ and consider the orbit $\poisson{D(g)v_A\otimes D(g)v_B\vert g\in G}$ where $D(g)$ is an unitary irreducible representation of $G$ acting in Alice and Bob space. The choice of $v_A\otimes v_B$ is crucial. It is assumed that it is done in such a way that the elements of the orbit are in one-to-one correspondence with the pairs $\poisson{\naw{A_i,\mu_i},\naw{B_j,\nu_j}}$ where $A_i$ ($B_j$) are Alice (Bob) observables while $\mu_i$ ($\nu_j$) number the eigenvectors of $A_i$ ($B_j$). Then the sum (\ref{b}) can be written as
\begin{equation}
S=\sum_{g\in G}\modu{\naw{D(g)v_A\otimes D(g)v_B,w}}^2
\end{equation}
or, introducing the operator
\begin{equation}
X\naw{v_A,v_B}=\sum_{g\in G}\naw{D(g)v_A\otimes D(g)v_B}\naw{D(g)v_A\otimes D(g)v_B}^+\label{c6}
\end{equation}
in the form
\begin{equation}
S=\naw{w,X\naw{v_A,v_B}w}.\label{ab1}
\end{equation}
In order to find an upper bound on $S$ we have to find the maximal value of the quadratic form on the right hand side of eq.~(\ref{ab1}) over the set of all normalized state vectors ($\Vert w\Vert=1$) describing the state of total system ($w$ is, in general, not of product form and describes entanglement);the corresponding $w$ describes the state of the whole system for which the upper bound on $S$ is attained. The state vectors of the total bipartite system transform under $G$ according to the representation $D\otimes D$ which, in general, is reducible. It decomposes into direct sum of irreducible representations $D^s$:
\begin{equation}
D\otimes D=\bigoplus\limits_s D^s.\label{ab3}
\end{equation}

Now, one can use the property that the maximal value of the quadratic form equals the maximal eigenvalue of the corresponding hermitean operator. Therefore, one has only to find the maximal eigenvalue of $X(v_A,v_B)$. This is greatly simplified by the fact that $X(v_A,v_B)$ commutes with all operators $D(g)\otimes D(g)$. Using Schur lemma one finds that it takes block-diagonal form in the basis in which the decomposition (\ref{ab3}) is explicit. Moreover, if it happens that any irreducible representations $D^s$ appears in (\ref{ab3}) with the multiplicity at most one, $X(v_A,v_B)$ is simply diagonal and reduces to the multiplicity of unit matrix in any irreducible subspace. Then its eigenvalues can be easily found \cite{Guney1} using orthogonality relations for group representations. They read
\begin{equation}
x_s=\frac{\modu{G}}{d_s}\Vert\naw{v_A\otimes v_B}_s\Vert^2\label{ed1}
\end{equation}
where $\modu{G}$ is the order of $G$, $d_s$ - the dimension of $D^s$ while $\naw{v_A\otimes v_B}_s$ - the projection of $v_A\otimes v_B$ onto the carrier space of $D^s$. In order to find an upper bound on $S$ we have to select the maximal eigenvalue $x_s$; therefore,
\begin{equation}
S\leq \max\limits_{s}\frac{\modu{G}}{d_s}\Vert\naw{v_A\otimes v_B}_s\Vert^2\equiv \frac{\modu{G}}{d_{s_0}}\Vert\naw{v_A\otimes v_B}_{s_0}\Vert^2.\label{ab4}
\end{equation}
Moreover, the bound is attained for any vector $w$ belonging to the subspace carrying the representation $D^{s_0}$. Let us note further that any vector belonging to $D^{s_0}$ can be written as a combination of product vectors with the coefficients being the relevant Clebsch-Gordan coefficients. Therefore, its degree of entanglement is determined on purely group-theoretical ground.

One can view the bound (\ref{ab4}) as Cirel'son - like bound in the sense that it determines the maximal value of the sum of probabilities corresponding to a fixed set of pairs of observables generated by group action.

Let us note that $O_{v_A}=\poisson{D(g)v_A\vert g\in G}$ ($O_{v_B}=\poisson{D(g)v_B\vert g\in G}$) defines an orbit of $G$ in Alice (Bob) space of states $\mathcal{H}$; the relevant orbit in the total space of states will be denoted by $O_{v_A\otimes v_B}=\poisson{D(g)v_A\otimes D(g)v_B\vert g\in G}$. The operator $X\naw{v_A,v_B}$ is uniquely defined by the choice of $O_{v_A\otimes v_B}$. It is easy to see that the operators $X$ corresponding to different orbits commute with each other. Therefore, they can be simultaneously diagonalized. By determining the maximal eigenvalue of the sum of $X$'s corresponding to the selected set of orbits $O_{v_A\otimes v_B}$ in $\mathcal{H}\otimes \mathcal{H}$ one finds various Cirel'son-like bounds on the relevant sums of probabilities.

\section{Group-theoretical framework for observables parametrization}

In order to apply group theory to find the upper bounds on relevant sums of probabilities one has to ascribe the elements of orbits $O_A (O_B)$ to the eigenvectors of some Alice (Bob) observables. This is a nontrivial task. Indeed, the observables are defined by their eigenvectors which provide the orthonormal bases in $\mathcal{H}$. Therefore, an orbit must consist of a number of such bases. In order to construct the appropriate orbits we will make some assumptions which, in particular, are valid for the examples of nonabelian groups studied so far in the literature. First, let us note that the dimension of  any irreducible over $\mathbb{C}$ representation $D$ of a given finite group $G$ divides the group order $\modu{G}$: $\frac{\modu{G}}{m}=k\in N, m=\dim D$ \cite{Simon}. It is also well known that $D$ can be put in unitary form, $D(g)\in U(m)$ (but, in general, $D(g)\notin SU(m)$). Assume that $G$ possesses a cyclic subgroup $H\subset G$ of order $m$, $H=\poisson{g^l\vert l=0,\ldots, m-1}$, $\naw{g^m=e}$. For $v\in \mathcal{H}$ the orbit $O_v$ is called regular if it consists of exactly $\modu{G}$ elements. Assume further that there exists a (normalized) vector $v\in \mathcal{H}$ such that: (a) $O_v$ is regular, and (b) $\naw{v,D(g^l)v}=0$ for $l=1,\ldots, m-1$. Let $0\leq p<l\leq m-1$; by unitarity and group property one has $\naw{D(g^p)v,D(g^l)v}=\naw{v, D(g^{l-p})v}=0$. Therefore, $\poisson{v, D(g)v,\ldots, D(g^{m-1})v}=\poisson{D(\widetilde{g})v\vert \widetilde{g}\in H}$ form an orthonormal basis in $\mathcal{H}$. It defines the spectral decomposition of some observable $A_1$. Consider now the set of left cosets of $G$ with respect to the subgroup $H$. Let $\poisson{g_1=e,g_2,\ldots, g_k}$ be any set of representatives in $G/H$. By unitarity of $D(g)$, the set $\poisson{D(g_\alpha g^l)v\vert l=0,1,\ldots,m-1}$ forms an orthonormal basis for any $\alpha=1,\ldots,k$. It defines the spectral decomposition of some observable $A_\alpha$. Now, any $\widetilde{g}\in G$ can be uniquely written as
\begin{equation}
\widetilde{g}=g_{\alpha}g^l,\quad \alpha=1,\ldots,k,\,\, l=0,\ldots,m-1.
\end{equation} 
Then 
\begin{equation}
D(\widetilde{g})v=D(g_\alpha)D(g^l)v\equiv v_{\alpha l}
\end{equation}
describes l-th eigenvector of $A_\alpha$ and 
\begin{equation}
O_v=\poisson{v_{\alpha l}\vert \alpha=1,\ldots,k\,\, l=0,\ldots,m-1}
\end{equation}
i.e. the orbit consists of $k$ orthonormal bases defining $k$ observables $A_\alpha$.

Let us note that changing the coset representatives amounts only to relabeling the eigenvectors of observables $A_\alpha$. This is irrelevant as far as we are not considering the eigenvalues. However, as we have noted above, everything can be formulated in terms of probabilities only so $g_\alpha$'s can be chosen at will.

Let us now choose any $\widetilde{g}\in G$ and consider the orbit generated by $D(\widetilde{g})v$. Writing $g_0=g_\alpha g^l$, one finds 
\begin{equation}
g_0\widetilde{g}= g_{\alpha_{\tilde{g}}}g^{l_{\tilde{g}}}
\end{equation} 
and it is straightforward to see that 
\begin{equation}
\naw{\alpha,l}\rightarrow \naw{\alpha, l}_{\widetilde{g}}\equiv\naw{\alpha_{\widetilde{g}},l_{\widetilde{g}}}
\end{equation}
is one-to-one mapping on the set of pairs $\naw{\alpha,k}$, $\alpha=1,\ldots,k$, $l=0,\ldots, m-1$, i.e. a permutation of such pairs.
In what follows we will be interested in the $G$-orbits in the total space of bipartite system which are of the form \cite{Guney1}$\div$\cite{Bolonek1}
\begin{equation}
O_v\naw{\widetilde{g}}=\poisson{D(g)v\otimes D(g)D(\widetilde{g})v\vert g\in G}.
\end{equation} 
They consists of the complete sets of eigenvectors of observables ascribed to both parties of the system. The operator $X\naw{v_A,v_B}$ (cf. eq. (\ref{c6})) with $v_A=v$, $v_B=D(\widetilde{g})v$ will be denoted by $X\naw{v,\widetilde{g}}$.

The following example provides the generalization of the constructions described in Ref. \cite{Guney1} ($G=S_3$) and \cite{Bolonek}, \cite{Bolonek1} ($G=S_4$). Let $G=S_n$ be the symmetric group. It acts in n-dimensional real space by permuting the components of any vector $x=\naw{x_1,x_2,\ldots,x_n}$. The real irreducible (also over $\mathbb{C}$) representation of $S_n$ obtained by imposing the invariant constraint $x_1+x_2+\ldots +x_n=0$ is called the standard representation of $S_n$; its dimension equals $n-1$. Let $H\subset G=S_n$ be the cyclic subroup generated by the cycle $g=\naw{12\ldots n-1}$. An orbit generated by $x$ ($\sum\limits_{i=1}^n x_i=0$) is regular iff $x_i\neq x_k$ for all $1\leq k\neq i\leq n$. The condition (b) above is equivalent to the set of equations
\begin{equation}
\begin{split}
& \sum_{i=1}^{n-1}x_ix_{i\oplus k}+x_n^2=0\\
& i\oplus k\equiv i+k(\text{mod}(n-1))\qquad k=1,\ldots,n-2.
\end{split}\label{c}
\end{equation} 
The number of independent equations (\ref{c}) equals $\com{\frac{n-1}{2}}$ while there are $n-1$ independent variables $x_1,x_2,\ldots, x_{n-1}$. Therefore, we can always find the solution generating regular  orbit. Alternatively, taking into account that $S_n$ is the symmetry group of regular $n-1$-simplex one could generalize the reasoning presented in Ref. \cite{Bolonek} for $S_4$ case.

\section{The one orbit case}
In the Refs \cite{Bolonek} and \cite{Bolonek1} the examples based on $S_3$ resp. $S_4$ were considered. It appeared that the Bell inequalities are not violated if only one orbit is involved. We shall show here that it is unlikely to find interesting examples of violation  based on one orbit only. Let us start with classical case. For any $\widetilde{g}\in G$ consider the orbit
\begin{equation}
O_v(\widetilde{g})=\poisson{v_{\alpha l}\otimes v_{\alpha_{\widetilde{g}}l_{\widetilde{g}}}\vert \alpha=1,\ldots,k,\, l=0,\ldots,m-1}.
\end{equation}
The relevant sum (\ref{a5}) takes the form
\begin{equation}
S=\sum_{\alpha, l}p\naw{\alpha,l;\alpha_{\widetilde{g}},l_{\widetilde{g}}};\label{c2}
\end{equation}
here $p\naw{\alpha,l;\alpha_{\widetilde{g}},l_{\widetilde{g}}}$ is the probability that Alice (Bob) observable $A_\alpha$ ($B_{\alpha_{\widetilde{g}}}$) acquires the value corresponding to the eigenstate $v_{\alpha l}$ ($v_{\alpha_{\widetilde{g}} l_{\widetilde{g}}}$). In order to find a classical upper bound on $S$ let us note that the right hand side of eq. (\ref{c2}) a priori consists of $\modu{G}$ terms. As we have noted above the upper bound can be always saturated by a probability distribution supported on a single joint configuration. Therefore, taking into account Alice observables we conclude that, for any $\alpha$, only one of the terms appearing on the right hand side and corresponding to this $\alpha$ can be nonvanishing (and, actually, equals one).

 As a result the sum (\ref{c2}) cannot exceed $\frac{\modu{G}}{m}=k$,
\begin{equation}
\sum_{\alpha,l}p\naw{\alpha,l;\alpha_{\widetilde{g}},l_{\widetilde{g}}}\leq k.\label{ab5}
\end{equation}
We shall show that this bound is saturated. To this end it is sufficient to show that one can select $k$ pairs $\naw{\alpha,l^{\naw{\alpha}}}$, $\alpha=1,\ldots,k$, such that $\alpha_{\widetilde{g}}\neq \alpha^{\prime}_{\widetilde{g}}$ for $\alpha \neq\alpha^\prime$. Then the probability distribution supported on the configuration $\poisson{\naw{\alpha,l^{(\alpha)}},\naw{\alpha_{\tilde{g}},l^{(\alpha)}_{\tilde{g}}}\vert \alpha=1,\ldots,k}$ saturates the bound.

In order to show that the set $\poisson{\naw{\alpha,l^{(\alpha)}}\vert \alpha=1,\ldots,k}$ exists consider a bipartite graph $\Gamma$ with bipartite sets $A$ and $B$ consisting of vertices representing the left cosets of $G$ with respect to $H$ ($\modu{A}=\modu{B}=k$). The vertices $\alpha\in A$ and $\beta\in B$ are adjacent iff there exists $0\leq l\leq m-1$ such that $\naw{\alpha,l}_{\tilde{g}}=\naw{\beta,l_{\tilde{g}}}$. For any subset $C\subset A$ let $N_\Gamma\naw{C}$ be a subset in $B$ adjacent to some elements of $C$. Now, $\naw{\alpha,l}\rightarrow\naw{\alpha_{\tilde{g}},l_{\tilde{g}}}$ is one-to-one. Let $\widetilde{C}\equiv\poisson{\naw{\alpha_{\tilde{g}},l_{\tilde{g}}}\vert\alpha\in C, l=0,\ldots,m-1}$; $\widetilde{C}$ consists of $m\modu{C}$ elements. On the other hand to any $\beta\in N_\Gamma(C)$ there correspond at most $m$ elements $\naw{\beta,l}\in\widetilde{C}$. Therefore, we conclude that
\begin{equation}
\modu{C}\leq\modu{N_\Gamma(C)}.
\end{equation}
By Hall theorem \cite{Cameron} there exists a matching between $A$ and $B$. Taking into account our definition of $\Gamma$ we find that the set $\poisson{\naw{\alpha,l^{(\alpha)}}\vert\alpha=1,\ldots,k}$ does exist. 

Let us remaind that, according to the discussion presented in Sec. II, the inequality (\ref{ab5}) provides the necessary condition for the relevant probabilities to be representable on classical level.

Let us now consider the quantum case. Assume that the $m$ dimensional irreducible (over $\mathbb{C}$) representation $D(g)$ is real. One can always take $D(g)\in O(m)$.
It should be expected that the more entangled the state is the more likely is the violation of Bell inequality. The tensor product $D\otimes D$ contains the trivial representation of $G$ which is spanned by 
\begin{equation}
\chi=\frac{1}{\sqrt{m}}\sum_{i=1}^m e_i\otimes e_i
\end{equation}
with $\poisson{e_i}_{i=1}^m$ being any orthonormal basis in the representation space of $D$. $\chi$ is maximally entangled, $\text{Tr}_B\rho(\chi)=\frac{1}{m}\mathbbm{1}_A$, $\text{Tr}_A\rho(\chi)=\frac{1}{m}\mathbbm{1}_B$. On the other hand, $\chi$ is an eigenvector of $X\naw{v_A,v_B}$. According to eq. (\ref{ed1}) the corresponding eigenvalue reads
\begin{equation}
\frac{\modu{G}}{1}\cdot\frac{1}{m}\naw{v_A,v_B}^2\leq\frac{\modu{G}}{m}=k
\end{equation}
so the quantum bound is not greater than the classical one. Therefore, if only one orbit is taken into account even the maximal entanglement is not sufficient to define the probabilities which cannot be represented on classical level. It would be interesting to extend this analysis to more general states and essentially complex representations. 

\section{Conclusions}
We have outlined the general construction of group orbits in representation space which determine the form of Bell inequalities. The main property of such an orbit is that it can be represented as a disjoint sum of a number of orthonormal bases, each defining the spectral decomposition of some observable. The assumption that a given orbit in $m$-dimensional space contains an orthonormal basis leads in general to $\binom{m}{2}$ equations on $m-1$ independent components of a vector generating the orbit. However, the number of equation can be reduced to at most $m-1$ if we assume that the group $G$ contains a cyclic subgroup $H$ of order $m$ which generates one orthonormal basis. Then all other bases are generated by the remaining group elements and can be put in one-to-one correspondence with the relevant left cosets $G/H$. Once such a construction is performed one can reduce the problem of finding an upper bound on the sum of classical probabilities to the group combinatorics. On the other hand the quantum upper bound (the counterpart of Cirel'son bound) is found using representation theory of finite groups.

 Assuming the orbit has been constructed according to the above rules we were able to find the actual form of Bell inequality. Its right hand side equals simply the number of elements in $G/H$. In order to show that it is the least upper bound we used the well-known Hall theorem.
 
 Finally, we pointed out that in the case of real representations (but irreducible over $\mathbb{C}$) the vector spanning trivial subrepresentation in the tensor product $D\otimes D$ does not lead, in the case of one orbit, to the violation of Bell inequality in spite of the fact that it is maximally entangled.
 
 Our discussion could be extended in various directions. First, it is interesting to find whether the Bell inequality can be violated in one orbit case if an (essentially) complex representation and arbitrary vectors in total space of states are considered.
 
 It would be also desirable to consider two (and more) orbits case. It seems rather obvious that the G\"uney-Hillery form of Bell inequality is completely determined by the combinatorics of the group $G$. In particular, in the case of two orbits $O_v(e)$ and $O_v(\widetilde{g})$ it should depend on the order of the element $\widetilde{g}$. We have checked that from this perspective the choice leading to the violation of Bell inequality for $G=S_3$, made in Ref. \cite{Guney1}, is essentially unique.
 
 The above problems will be addressed to in a subsequent publication.

\end{document}